 \documentclass[aps,prl,nofootinbib, nobibnotes,showpacs,notitlepage,reprint,superscriptaddress,floatfix]{revtex4-2}

 \usepackage{graphicx}
 \usepackage{hyperref}
 \usepackage{amsfonts}
 \usepackage{amsmath}
 \usepackage{amssymb}
 \usepackage{mathbbol}
 \usepackage{mathtools}
 \usepackage[nameinlink, capitalize]{cleveref}
 \usepackage{siunitx}
 \usepackage[sectionbib]{bibunits}
 \usepackage{braket}
 \defaultbibliographystyle{apsrev4-2} 
\defaultbibliography{silq}

\DeclareMathAlphabet{\stdmathbb}{U}{msb}{m}{n}

 \makeatletter
 \DeclareRobustCommand\ket[1]{%
  \@ifnextchar\bra{\k@t{#1}\!}{\k@t{#1}}%
 }
 \newcommand\k@t[1]{{|{#1}\rangle}}
 \makeatother

\renewcommand{\l}[1]{ \Ket{\mathbb{#1}} }

\newcommand{\ca}[0]{$^{40}$Ca$^{+}$}

\usepackage{etoolbox} % if not already loaded

\makeatletter
\newcounter{edfigure}
\renewcommand{\theedfigure}{\arabic{edfigure}}

\newenvironment{edfigure}[1][]{
  \begin{figure}[#1]
    \def\@captype{edfigure}%
    \def\fnum@edfigure{{Extended Data Fig. \theedfigure}}%
}{
  \end{figure}
}

\newenvironment{edfigure*}[1][]{
  \begin{figure*}[#1]
    \def\@captype{edfigure}%
    \def\fnum@edfigure{{Extended Data Fig. \theedfigure}}%
}{
  \end{figure*}
}
\makeatother

\makeatletter
\newcommand{\ext@edfigure}{Extended Data Fig.}
\makeatother

\crefname{edfigure}{Extended Data Fig.}{Extended Data Figs.}
\Crefname{edfigure}{Extended Data Fig.}{Extended Data Figs.}

\begin{document}
 
\begin{bibunit}[apsrev4-2]
 % \addbibresource{silq.bib}
 \title{Error correction of a logical qubit encoded in a single atomic ion}
 % \author{Kyle DeBry, Nadine Meister, Agustin Valdes Martinez, \\ Colin Bruzewicz, Other LL staff in some order, Isaac Chuang, \\ John Chiaverini}
 \author{Kyle DeBry}
 \email{debry@mit.edu}
 \affiliation{Massachusetts Institute of Technology, Cambridge, Massachusetts, USA}
 \affiliation{Lincoln Laboratory, Massachusetts Institute of Technology, Lexington, Massachusetts, USA}

 \author{Nadine Meister}
 \affiliation{California Institute of Technology, Pasadena, California, USA}
 % \affiliation{MIT Lincoln Laboratory}

 \author{Agustin Valdes Martinez}
 \affiliation{Massachusetts Institute of Technology, Cambridge, Massachusetts, USA}
 \affiliation{Lincoln Laboratory, Massachusetts Institute of Technology, Lexington, Massachusetts, USA}

 \author{Colin D. Bruzewicz}
 \affiliation{Lincoln Laboratory, Massachusetts Institute of Technology, Lexington, Massachusetts, USA}

 \author{Xiaoyang Shi}
 \affiliation{Massachusetts Institute of Technology, Cambridge, Massachusetts, USA}
 
 \author{David Reens}
 \affiliation{Lincoln Laboratory, Massachusetts Institute of Technology, Lexington, Massachusetts, USA}
 
 \author{Robert McConnell}
 \affiliation{Lincoln Laboratory, Massachusetts Institute of Technology, Lexington, Massachusetts, USA}

 \author{Isaac L. Chuang}
 \affiliation{Massachusetts Institute of Technology, Cambridge, Massachusetts, USA}

 \author{John Chiaverini}
 \affiliation{Massachusetts Institute of Technology, Cambridge, Massachusetts, USA}
 \affiliation{Lincoln Laboratory, Massachusetts Institute of Technology, Lexington, Massachusetts, USA}
 
 \date{March 18, 2025}

% Articles start with a fully referenced summary paragraph, ideally of no more than 200 words, which is separate from the main text and avoids numbers, abbreviations, acronyms or measurements unless essential. It is aimed at readers outside the discipline. This summary paragraph should be structured as follows: 2-3 sentences of basic-level introduction to the field; a brief account of the background and rationale of the work; a statement of the main conclusions (introduced by the phrase 'Here we show' or its equivalent); and finally, 2-3 sentences putting the main findings into general context so it is clear how the results described in the paper have moved the field forwards.
\begin{abstract}
    Quantum error correction (QEC) \cite{knill1997theory, gottesman2010introduction, nielsen2010quantum} is essential for quantum computers to perform useful algorithms, but large-scale fault-tolerant computation remains out of reach due to demanding requirements on operation fidelity and the number of controllable quantum bits (qubits). Traditional QEC schemes involve encoding each logical qubit into multiple physical qubits, resulting in a significant overhead in resources and complexity. Recent theoretical work has proposed a complementary approach of performing error correction at the single-particle level \cite{pirandola2008minimal} by exploiting additional available quantum states \cite{gross2021designing, chiesa2020molecular, jain_2024_ae_codes}, potentially reducing QEC overhead. However, this approach has not yet been demonstrated experimentally, partly due to the difficulty of performing error measurements and subsequent error correction with high fidelity. Here we experimentally demonstrate QEC in a single atomic ion, reducing errors by a factor of up to 2.2 and extending the qubit’s useful lifetime by up to 1.5 times compared to an unencoded qubit. The qubit is encoded into spin-cat logical states \cite{yu2025schrodinger, omanakuttan2024fault, gupta2024universaltransversalgates, lim2023fault, lim2024experimental}, and we develop a scheme for autonomous error correction \cite{aharanov_measurement_free, sarovar_measurement_free, reiter_measurement_free} that does not require mid-circuit measurements of an ancilla. Our work is applicable to a wide variety of finite-dimensional quantum systems \cite{lim2024experimental, yu2025schrodinger, champion2024multifrequencycontrolmeasurementspin72, roy2025synthetichighangularmomentum, Furey2024ae_molecules}, and such encodings may prove useful either as components of larger QEC codes \cite{omanakuttan2024fault, omanakuttan_2023_spin_concatenation, keppens2025quditvsqubitsimulated, campbell2014fault_tolerant_qudit} or when used alone in few-qubit devices such as quantum network nodes. 
\end{abstract}

\maketitle
% \end{document}

% \section{Main}

% Why is improving quantum error correction important?

% How might trapped ion systems provide new opportunities for improving QEC?
%
% How might QEC with higher dimensional physical systems such as trapped ions offer improved performance over state of art?
%
% Want to add discussion of net advantage, decrease redundancy

% How might qudit-based error correction confer an advantage for implementing useful QEC?
Quantum error correction is essential for protecting fragile quantum information against errors from environmental perturbations and imperfect control % by decreasing error rates and increasing coherence times 
\cite{nielsen2010quantum, gottesman2010introduction, knill1997theory}, but utility-scale fault-tolerant computation remains out of reach. The logical error rate of error-correcting codes decreases with an increasing number of physical qubits, but only if the error rates of the physical qubits are below the code's ``threshold'' error rate. Finite-dimensional quantum systems with more than two states, known as \textit{qudits}, present a promising alternative to traditional QEC codes, allowing error correction to be performed entirely within a single particle \cite{gross2021designing, chiesa2020molecular, lim2023fault}. These qudit codes can be designed around a particular physical system, and can be especially effective for biased errors where one particular error mechanism is dominant---for example, dephasing in atomic systems.
%Encodings in qudits may then be treated as the physical qubits in a larger error-correcting code, but to be useful they must keep error rates further below the larger code's threshold than an unencoded physical qubit would.
Encodings in qudits may then be treated as the physical qubits in a larger error-correcting code, but to be useful they must outperform an unencoded physical qubit at keeping error rates below the code's threshold.

% Tension:
% - Error correction operation
% - Picking the right code
% - Picking the right system

\begin{figure}[htbp]
    \centering
    \includegraphics[width=\linewidth]{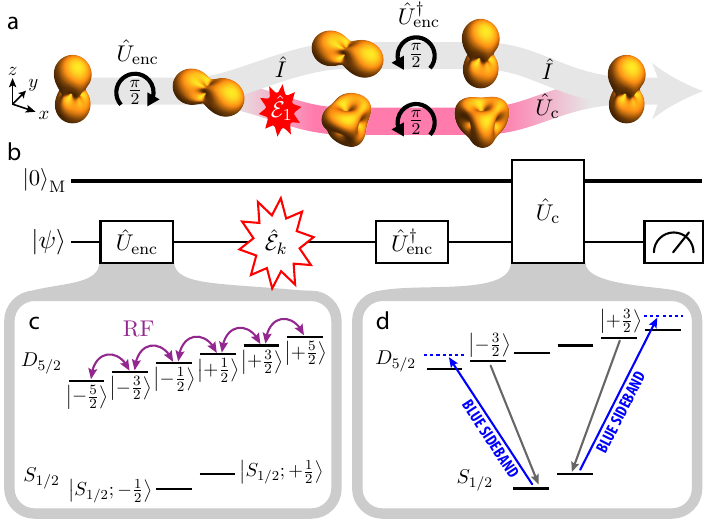}
    % How our error correction protocol works
    \caption{\textbf{Error correction protocol.} \textbf{a,} Evolution of the logical qubit in angular momentum space with and without an error. The angular momentum distribution $\left|\sum_{m_J} \Braket{m_J|\psi_\text{L}} Y_{5/2}^{m_J}(\theta, \phi)\right|$ is plotted, for polar angle $\theta$ and azimuthal angle $\phi$, at each step. The qubit is encoded (by a $\pi/2$ rotation about the $y$-axis), subjected to a first-order $\hat{J}_z$ error ($\hat{\mathcal{E}}_1$, leading to an orthogonal superposition of $\Ket{m_J}$ basis states) or not ($\hat{I}$), decoded (by a $-\pi/2$ rotation), and corrected with $\hat{U}_\text{c}$ (which acts as the identity if there is no error). \textbf{b,} Circuit diagram of the experiment, including the ion's motional harmonic oscillator mode labeled $\Ket{0}_\text{M}$ that is used for the autonomous error correction scheme and reset to $\Ket{0}_\text{M}$ between experiments. \textbf{c,}~Implementation of the encoding unitary $\hat{U}_\text{enc}$ as an RF $\pi/2$-pulse rotating the $D_{5/2}$ manifold. The RF Hamiltonian is depicted in terms of the energy level diagram of the \ca~ion (energy levels not to scale). \textbf{d,} Implementation of the correction unitary $\hat{U}_\text{c}$ via two carrier $\pi$-pulses and two blue sideband $\pi$-pulses of a \SI{729}{\nano\meter} laser that returns $\Ket{\pm\tfrac{3}{2}}$ population to $\Ket{\pm\tfrac{5}{2}}$.}
    \label{fig:circuit}
\end{figure}

Despite substantial theoretical progress, technical challenges have thus far prevented experimental demonstrations of qudit-based quantum error correction. Coherent control of the multiple levels comprising a qudit is difficult in many physical systems, limiting the ability to encode and decode logical states with high fidelity. Implementing the error correction operation is similarly challenging, as the typical approach of non-destructive syndrome measurements \cite{gottesman2010introduction} requires performing an entangling gate with an ancilla qubit, and qudit entangling gates have only recently been demonstrated \cite{hrmo2023native_qudit, ringbauer2022universal, goss2022high, luo2023experimental}. Finally, to enable net performance improvement, it is crucial that a 
code is available that maximizes correction of dominant errors while minimizing control errors.

In this work, we overcome these challenges and experimentally demonstrate a qudit-based QEC protocol that reduces overall error rates by a factor of up to 2.2 and extends the useful lifetime of a qubit encoded in the spin-5/2 manifold of a single \ca~ion by a factor of up to 1.5. There has been extensive pioneering theoretical work developing error correction codes tailored to the error mechanisms and symmetries of spin-$J$ systems \cite{albert2020molecule_codes, gross2021designing, omanakuttan_2023_spin_concatenation, omanakuttan2023_spin_gkp, omanakuttan2023_spin_clifford, gross2024hardware, jain_2024_ae_codes, kubischta_2025_permitation_invariant_codes, Furey2024ae_molecules, aydin_2025_ae_generalization}, including spin-cat logical qubits \cite{yu2025schrodinger, omanakuttan2024fault, gupta2024universaltransversalgates, lim2023fault, lim2024experimental}, which we implement here. The spin-cat codes (illustrated in \cref{fig:circuit}) effectively target dephasing, which is the dominant source of decoherence in \ca~ions. To perform high-fidelity encoding of the spin-cat logical states, we take advantage of the long coherence times, high-fidelity gates, and precise individual control that are characteristic of trapped-ion systems \cite{haffner2008quantum, bruzewicz2019trapped}, as well as their many accessible internal states.  Finally, we demonstrate a novel measurement-free error correction protocol \cite{butt_measurement_free, aharanov_measurement_free, reiter_measurement_free, veroni_measurement_free, heussen_measurement_free, sarovar_measurement_free, park2025_measurement_free_reinforcement_learning}, making use of the ion's ground-state-cooled motional mode as a resettable ancilla to efficiently perform the error recovery operation.

% Explain structure of paper?

\subsection{Hardware-efficient quantum error correction}
% What error correction code can produce a net improvement in coherence time?
% Define net advantage
% Expectation: linear vs quadratic in error
% Expectation: biased errors (dephasing)
% Set up expectations that match promised results

% Discuss autonomous error correction:
% How can we have scalable(?) error correction

\begin{figure}[htbp]
    \begin{flushleft}
    \end{flushleft}
    \centering
    \includegraphics[width=\linewidth]{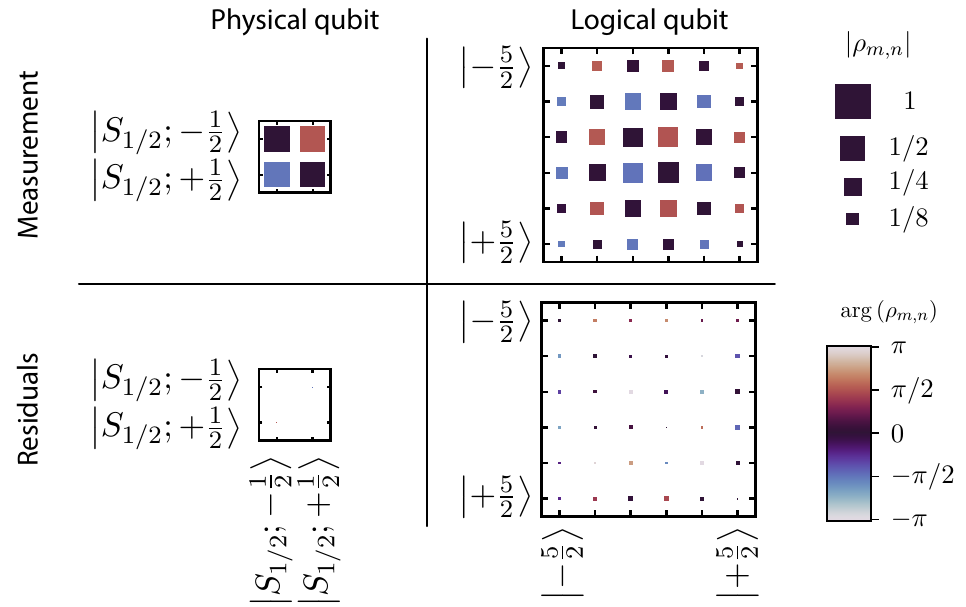}
    \caption{
    \textbf{High fidelity state preparation of physical and logical qubits.}
    Density matrices reconstructed with quantum state tomography (top row) and residuals compared to an ideal state (bottom row). The ground state physical qubit $\left(\Ket{S_{1/2};-\tfrac{1}{2}}-i\Ket{S_{1/2};+\tfrac{1}{2}}\right)/\sqrt{2}$ is shown in the left column with a fidelity of 0.9995, and the logical spin-cat qubit $\left( \l0 - i \l1 \right)/\sqrt{2}$ is shown in the right column with a fidelity of 0.997. For each matrix element $\rho_{m,n}$, the area is proportional to $\left|\rho_{m,n}\right|$, and the color denotes the phase $\arg\left(\rho_{m,n}\right)$, as shown on the right of the figure. The same scales are used for density matrices throughout the manuscript.}
    \label{fig:denmats}
\end{figure}

Our starting point is the definition of a clear performance metric suitable for making meaningful comparisons between the error rates of different types of qubits.
%Practical qudit-based error correction codes must keep the error rate low enough to be useful for longer than an unencoded qubit. 
%With a view towards the use of these logical qubits as inputs to a topological code with an error rate threshold below which errors are exponentially suppressed, 
%
Following standard practice \cite{nielsen2010quantum} in the pursuit of fault-tolerant quantum computation, it is natural to identify the ``useful lifetime'' $\tau(\epsilon)$ of a memory qubit as the duration for which its error compared to the ideal qubit state remains below a specified infidelity $\epsilon$ (see Methods). The improvement in the coherence time of a logical qubit $\tau_\text{L}$ over that of a physical qubit $\tau_\text{P}$ is then $\Lambda(\epsilon) \equiv \tau_{\text{L}}(\epsilon)/\tau_\text{P}(\epsilon)$. This figure of merit enables relevant comparisons between qubits, even when their fidelities decay according to different functional forms, particularly in the important small-$\epsilon$ regime.
%In particular, because quantum error correction of first-order errors eliminates the leading term in the infidelity as a function of the strength of the error, a logical qubit will necessarily display different behavior following one round of error correction, and using a single coherence time number to characterize a qubit is less useful. Typical $1/e$ coherence times do not reliably characterize the small-errors regime (errors on the order of 1\%) if the decay is not exponential, but this is the regime most relevant for quantum error correction.

The coherence time improvement $\Lambda(\epsilon)$ can be maximized by addressing the highly biased nature of errors in our experiment. Dephasing errors, often caused by fluctuating magnetic fields, are a key source of quantum memory error for atomic qubits and the dominant source of decoherence in our apparatus. Modeling our fluctuating magnetic field as being constant over a single experimental trial with shot-to-shot variation described by a Gaussian distribution of width $\sigma_B$, we define a dephasing parameter $\chi = \tfrac{1}{2}(\mu_\text{B} \sigma_B / \hbar)^2 t^2$ (\cite{chiesa2020molecular}, see Methods) such that the error $\epsilon$ of a qubit as a function of its interrogation time $t$ is proportional to $\chi$ to lowest order. Here, $\mu_\text{B}$ is the Bohr magneton and $\hbar$ is the reduced Planck constant. With error correction, the leading term in qubit error should be $\epsilon\sim\chi^2$, as opposed to $\epsilon\sim\chi$, when $\chi \ll 1$.

The elegantly simple spin-cat states of Refs.~\cite{yu2025schrodinger, omanakuttan2024fault, gupta2024universaltransversalgates} provide hardware-efficient \cite{gross2024hardware} logical qubits for \ca~ions giving a high degree of protection from dephasing errors caused by magnetic field fluctuations along the quantization axis. These states, also referred to as binomial states \cite{chiesa2020molecular, lim2024experimental}, are similar to other ``Schr\"{o}dinger's cat'' quantum states \cite{michael2016new_binomial, hu2019quantum_binomial}. They are formed from superpositions of spin-coherent states and can correct for up to $\lfloor{J-1/2}\rfloor$ orders of dephasing errors in a system with total angular momentum $J$. Spin-coherent states, which are SU(2) rotations of extremal angular momentum eigenstates, can be created straightforwardly in the long-lived $D_{5/2}$ manifold in \ca. We implement a rotated spin-cat code comprised of extremal eigenstates of $\hat{J}_x$:
\begin{align*}
    \l0 &= \frac{\Ket{-\tfrac{5}{2}}_x - \Ket{+\tfrac{5}{2}}_x}{\sqrt{2}};\quad
    \l1 = \frac{\Ket{-\tfrac{5}{2}}_x + \Ket{+\tfrac{5}{2}}_x}{\sqrt{2}},
\end{align*}
where $\Ket{m_J}_x = e^{i \hat{J}_y \pi/2} \Ket{m_J}$ refers to the $D_{5/2}$ manifold's eigenstate of $\hat{J}_x$ with eigenvalue $m_J$, and $\Ket{m_J}$ (without subscript) refers to the $m_J$ eigenstate of $\hat{J}_z$. Operators $\hat{J}_i$ are the $i$-axis projections of the total angular momentum. The effect of dephasing errors can be decomposed into a discrete set of error operators $\set{\hat{\mathcal{E}}_k} = \set{\hat{J}_z^k}$ (indexed by $k \in \stdmathbb{N}_0$) by following a standard procedure (see Methods), and the spin-cat codewords protect against $\hat{J}_z$ and $\hat{J}_z^2$ errors in a $J=5/2$ system.
%
% as error operators can be discretized following the standard procedures to satisfy the Knill-Laflamme conditions  
%

We create spin-cat logical superpositions by preparing a ground state superposition of $\Ket{S_{1/2};-\tfrac{1}{2}}$ and $\Ket{S_{1/2};+\tfrac{1}{2}}$, which is then coherently transferred to the $D_{5/2}$ manifold with two laser $\pi$-pulses, creating a superposition of $\Ket{\pm \tfrac{5}{2}}$. Then the encoding unitary operator $\hat{U}_\text{enc}$ is applied via a radio-frequency-driven SU(2) $\pi/2$-rotation of the $D_{5/2}$ manifold \cite{debry_mbp} as shown in \cref{fig:circuit}c, creating the desired logical state $\Ket{\psi_\text{L}} = \alpha \l0 + \beta \l1$. We prepare each of the six cardinal logical qubit states with fidelities $\ge0.996$; an example is shown in \cref{fig:denmats}.

% add sentence with takeaway message about theoretical predictions vs chi

\subsection{Measurement-free error correction protocol}

% need entropy dump, providing this is often hard

Achieving the predicted fidelity gain from the spin-cat codes requires removing entropy from the ion's internal state, but implementing this via ancilla readout is technically challenging and prone to two-qubit gate errors. Instead, we devise and implement a measurement-free, autonomous error correction scheme to correct first order ($\hat{\mathcal{E}}_1$) errors without the use of two-qubit gates or ancilla measurements \cite{butt_measurement_free, aharanov_measurement_free, reiter_measurement_free, veroni_measurement_free, heussen_measurement_free, sarovar_measurement_free, park2025_measurement_free_reinforcement_learning}. The logical decoding pulse $\hat{U}_\text{enc}^\dagger$ effectively converts $\hat{J}_z$ phase errors into $\hat{J}_x$ amplitude errors (see \cref{fig:spin-cat} in Methods), with errors leading to a mixed state. To first order in $\chi$, the state created by a first-order error $\Ket{\mathcal{E}_1} = \alpha \Ket{-\tfrac{3}{2}} + \beta \Ket{+\tfrac{3}{2}}$ occurs with probability $p_1$ proportional to $\chi$ and the desired decoded state $\Ket{\psi}=\alpha\Ket{-\tfrac{5}{2}} + \beta \Ket{+\tfrac{5}{2}}$ occurs with probability $p_0 \approx 1 - p_1$. The density operator describing the ion's internal state is thus
%with first order errors leading to a mixed state containing a coherent superposition of the states $\Ket{\pm (J-1)}$, labeled $\e0$ and $\e1$ as shown in \cref{fig:circuit}d, in addition to the error-free population in $\Ket{0} \equiv \Ket{-5/2}$ and $\Ket{1} \equiv \Ket{+5/2}$:
% \begin{align*}
%     \rho_\text{dec} & = p_0 \left(\alpha \Ket{0} + \beta \Ket{1} \right)\left(\alpha^* \bra{0} + \beta^* \bra{1}\right) \\
%     & + p_1 \left(\alpha \e0 + \beta \e1 \right)\left(\alpha^* \bra{\mathcal{E}_1 0} + \beta^* \bra{\mathcal{E}_1 1}\right) \\
%     & + \cdots
% \end{align*}
\begin{align*}
    \rho_\text{dec} & \approx p_0 \Ket{\psi}\Bra{\psi} + p_1 \Ket{\mathcal{E}_1}\Bra{\mathcal{E}_1} + O\left(\chi^2\right),
\end{align*}
where higher-order terms are omitted.
Correcting the errors amounts to performing a one-way operation that moves the populations in $\Ket{\mathcal{E}_1}$ back into the qubit subspace $\Set{\Ket{-\tfrac{5}{2}},\Ket{+\tfrac{5}{2}}}$ while preserving the coherence of the $\Ket{\mathcal{E}_1}$ superposition.

Using the motional state of the ion as a resettable ancilla, we implement the error recovery operator $\hat{U}_\text{c}$ by transferring the entropy associated with the error to the motional mode. Because $\hat{U}_\text{enc}^\dagger$ transforms the errors into distinct $\Ket{m_J}$ states, we can use spectroscopically-resolved laser pulses to implement $\hat{U}_\text{c}$. The population of the two error states ($\Ket{\pm \tfrac32}$) is transferred to the $\Ket{S_{1/2};\pm \tfrac12}$ ground states via two $\pi$-pulses of \SI{729}{\nano\meter} light. Then, two blue sideband (motion-adding) $\pi$-pulses are applied on the $\Ket{S_{1/2};\pm \tfrac12} \Ket{n}_\text{M} \leftrightarrow \Ket{\pm \tfrac52}\Ket{n+1}_\text{M}$ transition, where $\Ket{n}_\text{M}$ labels the motional state of the ion. This sequence is illustrated in \cref{fig:circuit}d, and the effect of $\hat{U}_\text{c}$ on the internal state is shown in \cref{fig:correction}b (without $\hat{U}_\text{c}$) and \cref{fig:correction}c (with correction) where the return of the population from $\Ket{\pm \tfrac32}$ to $\Ket{\pm \tfrac52}$ is observable. Provided the motion is initialized to the ground state, only the $S_{1/2} \rightarrow D_{5/2}$ direction of the sideband pulse is driven (and not the reverse), leaving the internal and motional state of the ion in
\begin{align*}
    \rho_\text{cor} \approx \left(\Ket{\psi}\bra{\psi}\right) \otimes \left( p_0 \Ket{0}_\text{M}\Bra{0}_\text{M} + p_1 \Ket{1}_\text{M}\Bra{1}_\text{M}\right),
\end{align*}
to leading order in the error strength (see Methods for a more detailed discussion). The internal state of the ion is then in the desired (pure) state, leaving no residual entanglement between the internal electronic state of the ion and its motion.

While this protocol is susceptible to errors from non-zero motional excitation, these can be reduced via sympathetic cooling \cite{kielpinski_cooling, blinov_cooling, barrett_cooling, home_cooling, allcock2021omg} and remaining errors can be detected and converted to erasure errors \cite{ohzeki_erasure_threshold, Wu2022_erasure, shi2024, kang_erasure_ions}. We implement such an erasure-conversion protocol (see Methods), marking up to $7\%$ of trials as erasures for the longest delay times, consistent with the measured motional heating rate. Aside from residual experimental control errors, which are expected to be insensitive to delay time, the internal state error of the ion will be reduced to $O(\chi^2)$. By using either an additional motional mode or mid-experiment recooling, both first- and second-order errors could be corrected by also transferring the $\Ket{\pm\tfrac12}$ populations to the $\Ket{\pm\tfrac52}$ states (see Methods).

% new theory prediction?  given this, reader should have no more surprises regarding theory predictions / expectations

\subsection{Improvement beyond break-even}

\begin{figure*}[htbp]
    \centering
    \includegraphics[width=0.9\linewidth]{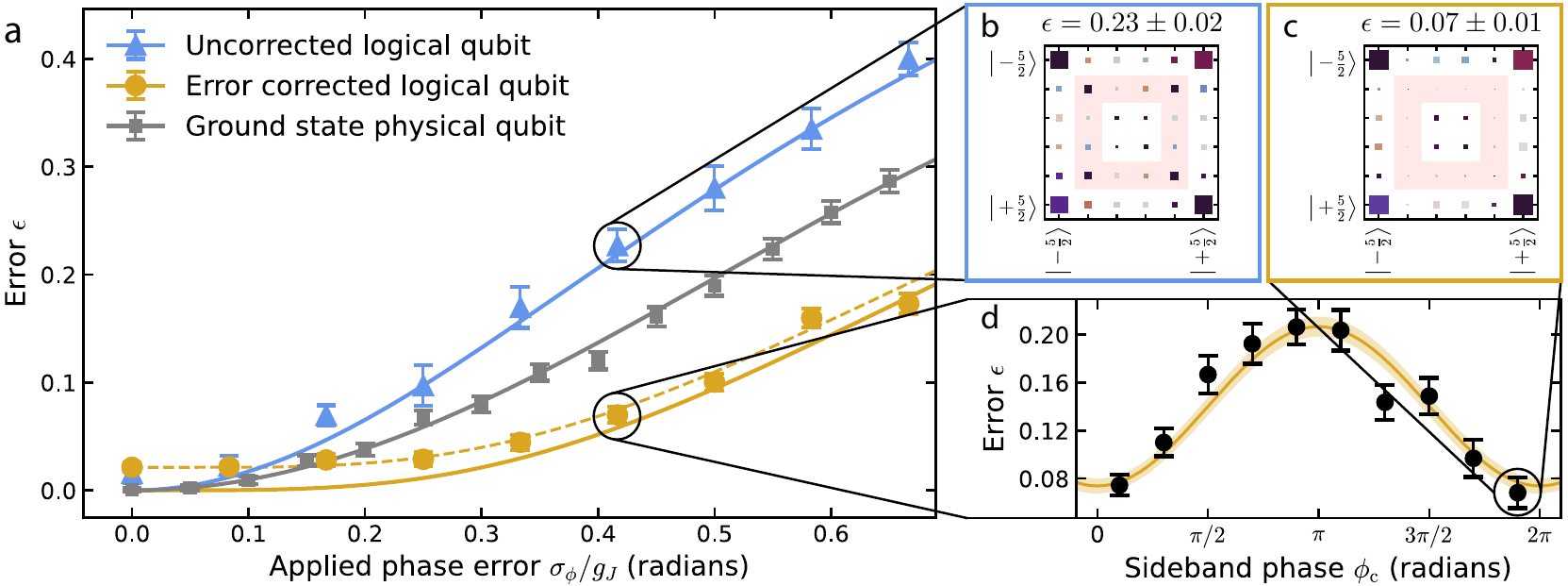}
    \caption{\textbf{Correction of first order $\hat{J}_z$ errors from applied $\hat{J}_z$ rotations.} \textbf{a,} Error of the uncorrected logical qubit, error-corrected logical qubit, and ground state physical qubit as a function of the applied phase noise strength $\sigma_\phi$, scaled by the appropriate Land\'{e} $g$-factor. Solid lines are theory curves with no free parameters, and the dashed gold line includes a measured $\sigma_\phi$-independent dephasing error contributing $0.022(4)$ to $\epsilon$ (Methods). \textbf{b,} Measured density matrix for the uncorrected logical qubit for $\sigma_\phi/g_J=5/12$, with first-order errors highlighted in pink. \textbf{c,} Measured density matrix for an error-corrected logical qubit for $\sigma_\phi/g_J=5/12$, demonstrating correction of first-order errors. \textbf{d,} Error of a logical qubit after error correction for different sideband phases $\phi_\text{c}$, with a fit used to extract the optimal error correction fidelity.}
    \label{fig:correction}
\end{figure*}

We experimentally evaluate error reduction under two conditions: first, using errors artificially generated from a simulated noise source, and second, with errors that are naturally occurring. When employing calibrated $\hat{J}_z$ rotations to simulate errors, we observe that the error-correction procedure accurately reproduces theoretical predictions and improves the qubit's fidelity.
%The error-correction procedure accurately reproduces theoretical predictions and improves the qubit's fidelity when calibrated $\hat{J}_z$ rotations are employed to simulate errors. 
After the logical qubit is prepared with $\hat{U}_\text{enc}$, an $e^{-i\hat{J}_z \phi}$ rotation error is applied via a radio-frequency (RF) pulse sequence, with $\phi$ drawn from a Gaussian distribution of width $\sigma_\phi/g_J$ where $g_J$ is the manifold's Landé $g$-factor, followed by decoding pulse $\hat{U}_\text{enc}^\dagger$. In \cref{fig:correction}, we plot the fidelity with and without the error correction operation $\hat{U}_\text{c}$ for a range of $\sigma_\phi/g_J$, as well as the fidelity of a ground state qubit prepared as $\left(\Ket{S_{1/2};-1/2} - i\Ket{S_{1/2};+1/2}\right)/\sqrt{2}$ (see \cref{fig:denmats}) subject to ground state rotation errors that would be produced by the same magnetic field variation. Solid lines are theoretical calculations with no free parameters. Fidelities are extracted from tomographic reconstructions (shown in \cref{fig:correction} and described in the Methods). The dashed gold curve includes a measured $\sigma_\phi$-independent dephasing error contributing $0.022(4)$ to $\epsilon$, arising from imperfections in the experimental implementation. %In the best case ($\sigma_\phi$ = 0.4), we measure a reduction in error rate for the encoded qubit of 3.8(7) times compared to the case where the logical qubit is prepared but the error correction operation $\hat{U}_\text{c}$ is not performed. 

\begin{figure*}
    \centering
    \includegraphics[width=1\linewidth]{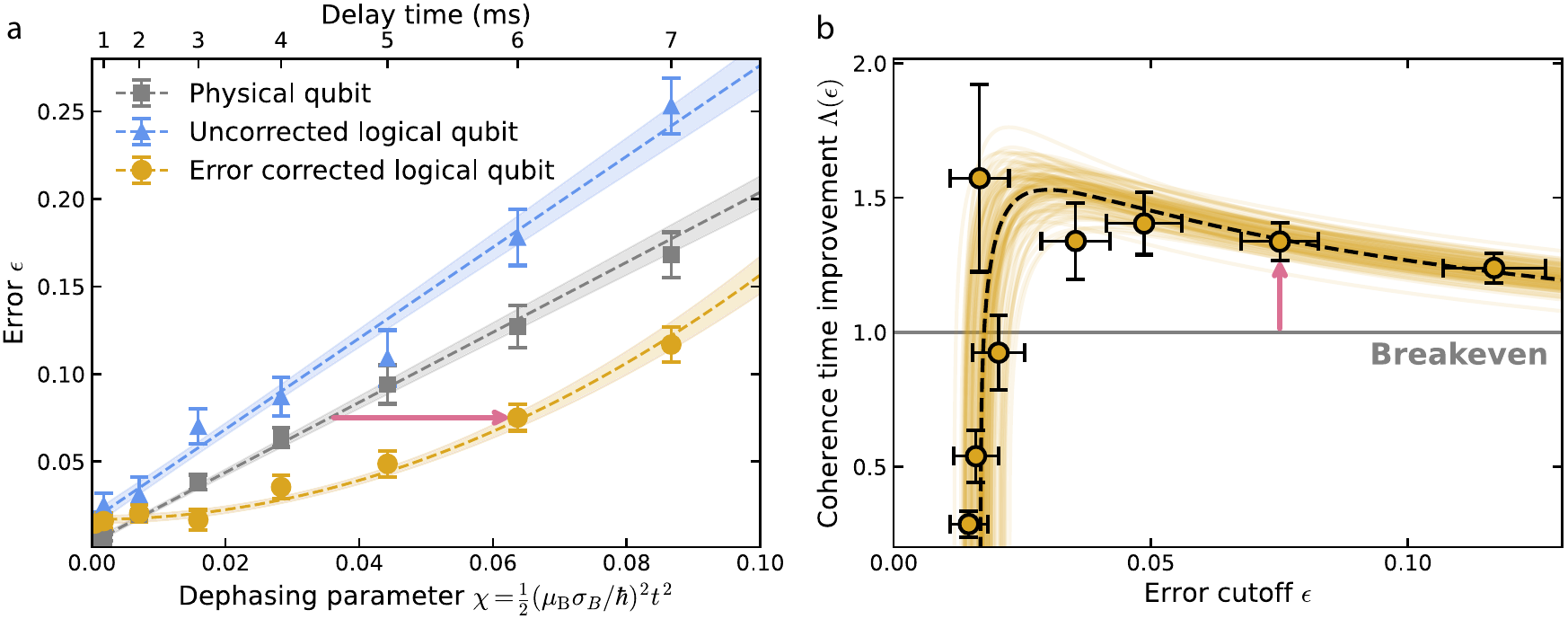}
    \caption{\textbf{Error correction extending lifetime beyond break-even under realistic environmental conditions.} \textbf{a,} Error rate of a ground-state physical qubit, an uncorrected logical qubit, and an error-corrected logical qubit as delay time increases under identical environmental conditions, plotted versus the dephasing parameter $\chi$ (lower axis) and delay time $t$ (upper axis). Dashed lines and shaded uncertainties are linear (quadratic) fits to the uncorrected (corrected) data. \textbf{b,} Increase in the useful lifetime of an error corrected qubit compared to the ground state physical qubit, defined as $\Lambda(\epsilon)$ in the text, for different error cutoffs $\epsilon$. To illustrate fit uncertainties, we draw 100 random realizations of the fit parameters from the best-fit covariance matrix and plot in pale yellow. The red arrows in \textbf{a} and \textbf{b} indicate the same logical qubit data point's improvement over the lifetime of the physical qubit.}
    \label{fig:breakeven}
\end{figure*}

When the noise is naturally occurring, we also observe the expected error reduction and achieve a coherence time improvement of $\Lambda(\epsilon) > 1$, indicating beyond-breakeven performance of the error-corrected qubit. \Cref{fig:breakeven} shows the measured error for logical and physical qubits (under conditions detailed in the Methods) as the delay time $t$ and dephasing parameter $\chi \propto t^2$ increase, as well as fits to each data series. The data demonstrate the expected change in scaling from $\epsilon \sim \chi$ without error correction to $\epsilon \sim \chi^2$ with error correction, and the error-corrected qubit has a factor of up to $2.17(16)$ lower error than the unencoded physical qubit. We use the $(t,\epsilon)$ coordinates of each data point to fit its useful lifetime $\tau(\epsilon)$, and the coherence time improvement $\Lambda(\epsilon)=\tau_\text{L}(\epsilon)/\tau_\text{P}(\epsilon)$ is plotted in \cref{fig:breakeven}b. We observe an improvement of up to $\Lambda(\epsilon)=1.53(7)$ at $\epsilon=0.030(5)$.

Imperfections in state preparation, error correction, and measurement lead to the nearly constant offset in the logical qubit error as a function of the physical error strength $\chi$. The leading error sources are detuning errors during the finite-duration pulse sequences ($\sim1\%$), state preparation and measurement error ($\sim0.4\%$), and fluctuating light shifts imparted by the \SI{729}{\nano\meter} laser pulses due to laser power fluctuations ($\sim0.3\%$). The fidelity of these operations can be improved significantly, as single-qubit microwave-based (laser-based) quantum logic gates have been demonstrated at the $10^{-7}$ level ($10^{-5}$ level) \cite{smith2024singlequbitgateserrors107, moses2023_quantinuum_racetrack}. Our demonstration does not yet reduce error rates below the $\approx1\%$ threshold reported for the surface code \cite{fowler2009high}, but our analysis indicates no insurmountable technical barriers to doing so.

\subsection*{Outlook}

This demonstration of single-particle error correction opens the door for the practical use of qudit codes to address dominant noise sources in quantum computers. Spin-based qudit codes are broadly applicable due to the ubiquity of spin systems in nature and the ability to simulate spins in other systems \cite{champion2024multifrequencycontrolmeasurementspin72, roy2025synthetichighangularmomentum}. These small codes can be treated as the physical qubits in a fully fault tolerant code, helping to bring the system's ``physical'' error rate below the larger code's error threshold. And, because the error correction procedure can be identical and autonomous on each qudit, only global pulses are required, minimizing technical overhead. Furthermore, qudit-specific versions of traditional codes are possible. There have been some theoretical studies of qudit-based surface codes, with suggestions of higher thresholds than the qubit-based surface code \cite{keppens2025quditvsqubitsimulated, campbell2014fault_tolerant_qudit, omanakuttan2024fault}, but further investigation is needed.

System-relevant codes are able to correct specific errors more efficiently than typical qubit-based codes. By focusing on dominant error mechanisms, rather than unstructured depolarizing errors, these codes can have an outsized impact on an error budget. For instance, by focusing on dephasing errors as we have done here, up to two orders of $\hat{J}_z$ errors can be fully corrected in the six-dimensional $D_{5/2}$ manifold of \ca~ions. This is akin to a code of distance~5, despite the Hilbert space being smaller than that of a three-qubit phase flip code of distance~3~\cite{gross2024hardware, gottesman2010introduction}. In fact, spin-cat codes are ``perfect'' codes for any half-integer spin, saturating the quantum Hamming bound. Of particular interest, second-order ($\hat{J}_z^2$) errors arising from unwanted off-resonant light shifts, which are common in atomic experiments, can be corrected.

We posit that even this initial demonstration of a single-particle code could provide near-term benefit in small quantum systems. For example, finite coherence times are often among the largest error sources in implementations of photon-mediated entanglement generation in ions, atoms, and solid-state systems \cite{oreilly2024fast_entanglement_138ba, stephenson2020high_entanglement_88sr, zhou2024memory_entanglement_neutrals, uysal2024spinphotonentanglement_er3plus, kalb2017solid_state_entanglement}. This is the case because the requirement of a simple energy level structure in order to generate entangled photons often conflicts with the desire for field-insensitive qubit states only found in atoms with hyperfine structure. In real-world scenarios, this problem is exacerbated by the long memory times required due to slow entanglement rates. Recent trapped-ion experiments even perform a two-qubit gate after entanglement is generated to transfer the state to another ion species with a longer coherence time \cite{main2025distributed_entanglement_swap}. All of the physical platforms mentioned above are candidates to implement spin-cat codes, and could therefore benefit from increased coherence times without the significant technical overhead needed to swap the entanglement to another type of qubit.

We hope that the realization of single-atom error correction spurs the further development of small, practical codes. Important error mechanisms to target include $T_1$ decay in superconducting systems, Rydberg-state decay in neutral-atom experiments, and entangling-gate errors broadly. Reducing such errors before concatenating in a larger code may lead to a significantly lower resource overhead for practical quantum information processing.

% \printbibliography
%  \bibliographystyle{apsrev4-2}
% \bibliography{silq}

\putbib
\end{bibunit}

\clearpage

% \end{document}

\section{Methods}\label{sec:methods}

\begin{bibunit}

\subsection{Theory} 

Below, we define the explicit error operators describing noise processes in our system, leading to the specification of logical codewords employed to satisfy the quantum error correction criteria. We then derive the expected performance of these codewords in the presence of noise, including certain experimental imperfections.

% Add pulse sequence diagram

\subsubsection{Error operators}

The error operators $\set{\hat{\mathcal{E}}_k}$ allow us to turn a continuous process ($e^{-i \hat J_z \phi}$ rotations for $\phi \in \stdmathbb{R}$) into a discrete set that can be dealt with using a corresponding set of error recovery operators. They must accurately capture the effects of the error mechanism, which is magnetic field fluctuations in our case. In the presence of a magnetic field $B$, the Zeeman Hamiltonian for the $D_{5/2}$ manifold is
\begin{align}
    \hat{H}/\hbar = g_{J} \mu_B B \hat{J}_z,
\end{align}
where $\hbar$ is the reduced Planck constant, $g_{J} \approx 6/5$ is the Landé $g$-factor for the $D_{5/2}$ manifold with $J=5/2$, $\mu_B$ is the Bohr magneton, and $\hat{J}_z$ is the $z$ projection of the total angular momentum operator. Information can be encoded in the six states labeled $\Ket{D_{5/2}; m_J}$, and if the magnetic field is known exactly, the fidelity of encoded quantum information is limited only by spontaneous decay (the lifetime of the $D_{5/2}$ states is \SI{1.2}{\second}). Any deviation or uncertainty in the magnetic field, however, applies rotations of the form $e^{-i \phi \hat{J}_z}$, with $\phi = \int \frac{g_J \mu_\text{B}}{\hbar} \delta B dt$ being the integrated phase arising from magnetic field fluctuations $\delta B$. We can convert this continuous space of errors into a discrete set by expanding the action of magnetic field errors $e^{-i \hat{J}_z \phi}$ into a set of Kraus operators $\set{\hat{\mathcal{E}}_k} = \set{\hat{I}, \hat{J}_z, \hat{J}_z^2, \ldots }$ that can be detected and corrected with an appropriate choice of codewords \cite{chiesa2020molecular, nielsen2010quantum}. Because $\hat{J}_z^2 \neq \hat{I}$ for $J>1/2$, there are multiple orders of $\hat{J}_z$ error to consider, unlike the qubit case where $\hat{\sigma}_z^2=\hat{I}$ for the Pauli matrices $\set{\hat{\sigma}_i}$.

To demonstrate the error correction protocol in a controlled fashion, we first implement the error operators $\hat{\mathcal{E}}_k$ using calibrated RF rotations. We emulate expected real-world errors, where unknown rotations $e^{-i \phi \hat{J}_z}$ occur with rotation angles given by a Gaussian random variable $\phi$ with standard deviation $\sigma_\phi$. On the $i$th shot of the experiment, we draw a random number $\phi_i$ from such a distribution, and apply the rotation sequence
\begin{align}
    U_\text{err}(\phi_i) = e^{i \hat{J}_y \frac{\pi}{2}} e^{-i \hat{J_x} \phi_i} e^{-i \hat{J}_y \frac{\pi}{2}},
\end{align}
which implements an overall rotation $e^{-i \phi_i \hat{J}_z}$. Averaging over many shots of the experiment, this error operation modifies the initial density matrix $\rho$ according to 
\begin{align}
    \rho' = \int_{-\infty}^{\infty} e^{-i \phi \hat{J}_z} \rho e^{i \phi \hat{J}_z} \frac{e^{-\phi^2/2\sigma_\phi^2}}{\sigma_\phi \sqrt{2\pi}} d\phi.
\end{align}

Integrating this, we find that the density matrix element $\rho_{m,n}$ is scaled down by a factor of $e^{-\frac{\sigma_\phi^2}{2}(m-n)^2}$, with the coherences decaying quadratically with distance from the diagonal, and quadratically in $\sigma_\phi$. In physical terms, $\sigma_\phi$ represents the product of the magnitude of the magnetic field fluctuations $\sigma_B$, the sensitivity of the states to magnetic field changes $g_J \mu_B$, and the delay time $t$ as $\sigma_\phi = \sigma_B g_J \mu_\text{B} t$. We similarly define the dephasing parameter $\chi$ as $(\sigma_B \mu_\text{B} t)^2/2$ such that the density matrix elements decay by $e^{-g_J^2 \chi (m-n)}$ and correcting first-order errors reduces the leading error term from $\epsilon \sim \chi$ to $\epsilon \sim \chi^2$.

 The Hilbert space of the $D_{5/2}$ manifold permits correction of not only first-order errors but also second-order $\hat{J}_z^2$ errors $\hat{\mathcal{E}}_2$ (light red in \cref{fig:spin-cat}). Correcting these would change the leading order error term from $\epsilon \sim \chi^2$ to $\epsilon \sim \chi^3$. Higher-dimensional spins would permit correction of even more error terms, including the non-commuting set $\set{\hat{J}_x, \hat{J}_y, \hat{J}_z}$. Second order errors could be corrected with minimal changes to the experimental apparatus by using an additional motional-mode ancilla in the measurement-free error correction scheme. In addition to the axial mode used in this work, a single ion's two radial modes can also be used to correct errors, and an $N$-ion chain has $3N$ motional modes. By orienting the \SI{729}{\nano\meter} laser to have a projection along all three motional axes, all of these modes can be used to store entropy and, by recooling to the ground state, remove it from the system.

\subsubsection{Logical codewords}
\begin{edfigure}[htbp]
    \centering
    \includegraphics[width=0.75\linewidth]{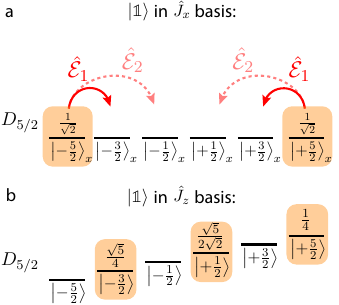}
    \caption{\textbf{Spin-cat codewords and the effects of error operators.} The logical $\l1$ codeword in the $\hat{J}_x$ and $\hat{J}_z$ bases. \textbf{a,} Spin-cat $\l1$ codeword in eigenbasis of $\hat{J}_x$, showing the action of error operators $\hat{\mathcal{E}}_1 = \hat{J}_z$ and $\hat{\mathcal{E}}_2 = \hat{J}_z^2$ in this basis. Support of the $\l1$ codeword is highlighted in orange. \textbf{b,} The $\l1$ codeword in the $\hat{J}_z$ basis, with the coefficients of the $\Ket{m_J}$ basis states written above each state. Support of $\l1$ in this basis is highlighted in orange.}
    \label{fig:spin-cat}
\end{edfigure}

The rotated spin-cat logical codewords used in this experiment convert difficult-to-correct phase errors into amplitude errors that can be corrected via a one-way operation performed with state-selective laser pulses. To illustrate the error-correction properties of the spin-cat codewords $\l0$ and $\l1$, we can first consider the action of $\hat{J}_x$ rotation errors on non-rotated spin-cat states, given by error operators $\hat{\mathcal{E}}_k^{(x)} \propto \hat{J}_x^k$. The action of $\hat{\mathcal{E}}_1^{(x)}=\hat{J}_x = \left(\hat{J}_+ + \hat{J}_- \right)/2$ on a superposition $\alpha \Ket{-J} + \beta\Ket{+J}$ is to map it to $\alpha \Ket{-J+1} + \beta \Ket{+J-1}$, as shown in \cref{fig:spin-cat}. The operators $\hat{J}_+$ and $\hat{J}_-$ are the angular momentum raising and lowering operators, respectively. A second order error $\hat{\mathcal{E}}_2^{(x)}$ will similarly shift the population by $\pm 2$ units of angular momentum towards the center of the manifold due to the $\hat{J}_\pm^2$ components of $\hat{J}_x^2$. Both of these are correctable errors in a spin-$5/2$ system via a recovery operator that returns the population from $\Ket{\pm(J-1)}\mapsto \Ket{\pm J}$ and $\Ket{\pm(J-2)} \mapsto \Ket{\pm J}$, while higher-order errors are not correctable.

We perform a change of basis in order to correct for the far more prevalent case of phase errors. By applying an SU(2) rotation of $\pi/2$ about the $y$-axis, we can convert $\hat{J}_z$-type phase errors into $\hat{J}_x$-type amplitude errors:
\begin{align}
    e^{-i \hat{J}_y \frac{\pi}{2}} \hat{J}_z e^{i \hat{J}_y \frac{\pi}{2}} = \hat{J}_x,
\end{align}
and these can then be corrected as described above. This amounts to encoding into logical states $\l0 = e^{i \hat{J}_y \pi/2}(\Ket{-J} - \Ket{J})/\sqrt{2}$ and $\l1 = e^{i \hat{J}_y \pi/2}(\Ket{-J} + \Ket{J})/\sqrt{2}$ (see \cref{fig:spin-cat} for examples of the logical states in the $\hat{J}_x$ and $\hat{J}_z$ basis).

\subsubsection{Error correction conditions}
% This section should discuss the code words we have chosen and show that they satisfy the error correction conditions.

Inspired by the binomial and Schr\"{o}dinger's cat codewords \cite{michael2016new_binomial, mirrahimi2014dynamically_cat}, we encoded our logical states using the codewords suggested by \cite{chiesa2020molecular, omanakuttan2024fault} for a $D_{5/2}$ system. The states in the $\hat{J}_z$ eigenbasis are
\begin{align}
    \ket{\mathbb{0}} &= \frac14 \Ket{-\frac52} + \frac12 \sqrt{\frac52} \Ket{-\frac12}+ \frac{\sqrt5}{4} \Ket{+\frac32},\\
    \Ket{\mathbb{1}} &= \frac{\sqrt5}{4}\Ket{-\frac32} + \frac12 \sqrt{\frac52} \Ket{+\frac12} + \frac14 \Ket{+\frac52}.
\end{align}

Error correction is possible if the Knill-laflamme error conditions \cite{knill1997theory} are satisfied. Namely, for any pair of error operators $\hat{\mathcal{E}}_j, \, \hat{\mathcal{E}}_k $ belonging to the error set $\mathcal{E} = \mathrm{span} \{\hat I, \hat J_z, \hat J_z^2\}$, the conditions %is the I necessary here? I think yes
\begin{align}
    \Braket{\mathbb{0} | \hat{\mathcal{E}}_j^\dagger \hat{\mathcal{E}}_k  | \mathbb{0}} &= \Braket{\mathbb{1} | \hat{\mathcal{E}}_j^\dagger \hat{\mathcal{E}}_k | \mathbb{1}}, \label{eq:ec-same-distinguish} \\
    \Braket{\mathbb{0} | \hat{\mathcal{E}}_j^\dagger \hat{\mathcal{E}}_k | \mathbb{1}} &= 0 \label{eq:ec-01-distinguish}
\end{align}
must hold.

The second condition $\braket{\mathbb{0} | \hat{\mathcal{E}}_j^\dagger \hat{\mathcal{E}}_k | \mathbb{1} } = 0$ is always satisfied because $\hat J_z$ and $\hat J_z^2$ do not change the magnetic quantum number $m_J$; they only affect its relative phase. 
Because $\l0$ and $\l1$ initially have no overlap in basis states $\ket{m_J}$, %again, states is not the right word...
applying $\hat{\mathcal{E}}_j^\dagger \hat{\mathcal{E}}_k $ cannot induce an overlap between them.

For the first condition (\cref{eq:ec-same-distinguish}), recall that $\hat J_z \ket{m_J} = m_J \ket{m_J}$ and $\hat J_z^2 \ket{m_J} = m_J^2 \ket{m_J}$, and thus we can check all pairs of $\hat I, \hat J_z, \hat J_z^2$ for $\hat{\mathcal{E}}_j, \, \hat{\mathcal{E}}_k$:
\begin{align*}
    \Braket{\mathbb{0} | \hat I^\dagger \hat J_z | \mathbb{0}} &= 0 = \Braket{\mathbb{1} | \hat I^\dagger \hat J_z | \mathbb{1}} \\
    \Braket{\mathbb{0} | \hat{I}^\dagger \hat{J}_z^2 | \mathbb{0}} &= \frac{5}{4}  = \Braket{\mathbb{1} | \hat{I}^\dagger \hat{J}_z^2 | \mathbb{1}} \\
    \Braket{\mathbb{0} | \hat{J}_z^\dagger \hat{J}_z^2 | \mathbb{0}} &= 0 = \Braket{\mathbb{1} | \hat{J}_z^\dagger \hat{J}_z^2 | \mathbb{1}} \\
    \Braket{\mathbb{0} | \hat{I}^\dagger \hat{I} | \mathbb{0}} &= 1 = \Braket{\mathbb{1} | \hat{I}^\dagger \hat{I} | \mathbb{1}} \\
    \Braket{\mathbb{0} | \hat{J}_z^\dagger \hat{J}_z | \mathbb{0}} &= \frac{5}{4}= \Braket{\mathbb{1} | \hat{J}_z^\dagger \hat{J}_z | \mathbb{1}} \\
    \Braket{\mathbb{0} | {\left({\hat{J}_z^\dagger}\right)}^2 \hat{J}_z^2 | \mathbb{0}} &= \frac{65}{16} = \Braket{\mathbb{1} | \left({\hat{J}_z^\dagger}\right)^2 \hat{J}_z^2 | \mathbb{1}}.
\end{align*}

%%%%%%%%%%%%%%
\begin{widetext}
\subsubsection{Predicted fidelities}
We analytically predict the performance of these states under $\hat{J}_z$ errors with and without error correction, and these models show good agreement with the experimental results. We use the following definition of the fidelity of a density matrix $\rho$ compared to the density matrix for the ideal (error-free) state $\rho_\text{ideal}$ \cite{Jozsa_fidelity}:
\begin{align}
    \mathcal{F}(\rho,\rho_{\mathrm{ideal}})=\left(\operatorname{Tr}\sqrt{\sqrt{\rho}\,\rho_{\mathrm{ideal}}\,\sqrt{\rho}}\right)^2, \label{eq:fidelity-def}
\end{align}
and thus define the error as $\epsilon = 1-\mathcal{F}(\rho,\rho_{\mathrm{ideal}})$. Recalling that the action of a Gaussian-distributed dephasing channel with width $\sigma_\phi$ on an input density matrix $\rho$ is
\begin{align}
    \left[\mathcal{E}(\rho)\right]_{m,n} = e^{-\frac{1}{2}\sigma_\phi^2\,(m-n)^2}\,\rho_{m,n},
\end{align}
we can calculate explicitly the fidelity of various input states by using the appropriate input density matrix and value of $\sigma_\phi$. Using $\chi=(\sigma_B \mu_\text{B} t)^2/2=\sigma_\phi^2/(2g_J^2)$ (adapted from the parameter $\lambda=\sigma^2/2$ defined in \cite{chiesa2020molecular}), and inserting the density matrix for a physical ground state qubit $\Ket{\psi_\text{P}}=\left(\Ket{S_{1/2};-1/2} + i\Ket{S_{1/2};+1/2}\right)/\sqrt{2}$, we see the expected fidelity
\begin{align}
    \mathcal{F}(\rho_\text{P},\rho_\text{ideal})=\frac{1+e^{-g_J^2 \chi}}{2}, \label{eq:phys-fidelity}
\end{align}
which gives a lowest-order error of $\epsilon \approx g_J^2 \chi/2 + O(\chi^2)$. Doing the same for the logical qubit state $\Ket{\psi_L}=\left(\Ket{-\tfrac{5}{2}}_x-i\Ket{+\tfrac{5}{2}}_x\right)/\sqrt{2}$ yields
\begin{align}
\mathcal{F}(\rho_\text{enc},\rho_\text{ideal}) =\frac{1}{512} e^{-25 g_J^2 \chi} \left( 1 + e^{9 g_J^2 \chi} \left( 10 + 3 e^{7 g_J^2 \chi} \left(15 + 40 e^{5 g_J^2 \chi} + 70 e^{8 g_J^2 \chi} + 42 e^{9 g_J^2 \chi}\right)\right)\right), \label{eq:enc-fidelity}
\end{align}
with lowest-order error $\epsilon \approx 5g_J^2 \chi/2 + O(\chi^2)$.

We can also use these calculated density matrices to compute the fidelity after error correction. To compute the effects of first-order error correction, we can take the density matrix of the encoded logical state after the error has been applied, and then act on the density matrix with the decoding operation $\hat{U}_\text{enc}^\dagger$. Assuming the error correction operator $\hat{U}_\text{c}$ moves the populations in the $\Ket{\pm\tfrac{3}{2}}$ states to the $\Ket{\pm\tfrac52}$ states without introducing any error, we can calculate the fidelity increase from first-order error correction by calculating the fidelity of the decoded density matrix with respect to the first-order error state $\Ket{\mathcal{E}_1}=\left(\Ket{-\tfrac{3}{2}}-i\Ket{+\tfrac{3}{2}}\right)/\sqrt{2}$. Adding this to the uncorrected fidelity, we get a total fidelity after first-order error correction ($\rho_\text{QEC}$) of
\begin{align}
    \mathcal{F}(\rho_\text{QEC},\rho_\text{ideal}) = \frac{1}{128} e^{-25 g_J^2 \chi} \left( -1 + e^{9 g_J^2 \chi} \left( -5 + e^{7 g_J^2 \chi} \left( -5 + 20 e^{5 g_J^2 \chi} + 70 e^{8 g_J^2 \chi} + 49 e^{9 g_J^2 \chi} \right) \right) \right). \label{eq:qec-fidelity}
\end{align}
This has a leading-order error of $\epsilon\approx15 g_J^4\chi^2/2 + O(\chi^3)$, and is now quadratic in $\chi$ rather than linear, as expected.

Finally, we can model the effects of finite error introduced by experimental imperfections. We model the effects of dephasing errors after the logical state has been decoded by $\hat{U}_\text{enc}^\dagger$, which could be induced by light shifts from the \SI{729}{\nano\meter} pulses used in $\hat{U}_\text{c}$ or detuning errors during the duration of $\hat{U}_\text{c}$. Assuming the dephasing errors are given by an independent Gaussian distribution of width $\delta$, the fidelity is
% \begin{align}
% &\mathcal{F}(\rho_\text{QEC},\rho_\text{ideal}) = \\
%     &\frac{1}{512} e^{-\delta - 25 g_J^2 \chi} \left( -1 + 5 e^{\delta} + 20 e^{\delta + 9 g_J^2 \chi} - 80 e^{\delta + 21 g_J^2 \chi} + 316 e^{\delta + 25 g_J^2 \chi} - 70 e^{24 g_J^2 \chi} \left(3 + e^{\delta}\right) + 5 e^{16 g_J^2 \chi} \left(-9 + 13 e^{\delta}\right) \right). \label{eq:qec-error-fidelity}
% \end{align}
\begin{equation}
\begin{split}
\mathcal{F}(\rho_\text{QEC},\rho_\text{ideal}) = \frac{1}{512} e^{-2 \delta^2 - 25 g_J^2 \chi} \Bigl( & -1 + 5 e^{2 \delta^2} + 20 e^{2 \delta^2 + 9 g_J^2 \chi} \\
& - 80 e^{2 \delta^2 + 21 g_J^2 \chi} + 316 e^{2 \delta^2 + 25 g_J^2 \chi} \\
& - 70 e^{24 g_J^2 \chi} \left(3 + e^{2 \delta^2}\right) + 5 e^{16 g_J^2 \chi} \left(-9 + 13 e^{2 \delta^2}\right) \Bigr). \label{eq:qec-error-fidelity}
\end{split}
\end{equation}

These fidelity calculations are used for analysis of the experimental data in the main text. In \cref{fig:correction}, the physical qubit theory curve is given by \cref{eq:phys-fidelity} with $g_J=2$ for the ground state. For the logical qubits in the $D_{5/2}$ manifold, $g_J=6/5$, and the encoded (but not corrected) qubit theory curve in \cref{fig:correction} is from \cref{eq:enc-fidelity} and the error corrected qubit theory curve is from \cref{eq:qec-fidelity}. The dashed gold curve is from \cref{eq:qec-error-fidelity}, with the value of $\delta$ extracted from a measurement of the fidelity with $\hat{U}_\text{c}$ applied in the absence of any introduced error. Similarly, in \cref{fig:breakeven}, the data are fit to simple polynomials with the expected forms from lowest-order Taylor expansions of the same curves. In particular, we expect the physical ground state qubit (with $g_J=2$) to follow $\epsilon \approx 2\chi+O(\chi^2)$, the encoded but not corrected logical qubit (with $g_J=6/5$) to follow $\epsilon \approx 18\chi/5+O(\chi^2)$, and the error-corrected logical qubit to follow $\epsilon\approx 1944/125\chi^2+O(\chi^3)$. In \cref{fig:breakeven}, the coefficients are fit parameters, and each fit also includes a constant ($\chi$-independent) error term to account for experimental imperfections. The coefficients from the ground state physical qubit is used to compute $\sigma_B = \SI{0.68(7)}{\nano\tesla}$ and $\chi$ in \cref{fig:breakeven}.

\end{widetext}

%%%%%%%%%%%%%%%%%%%%%%%%%%%%%%%%%%%%%%%%%%%%%%%%
\subsection{Experiment}

Below, we briefly describe our experimental pulse sequence and apparatus, elaborate on the calibration procedures employed, and specify the methods used for density matrix and fidelity measurements. We then provide procedures used for evaluation of the error correction demonstration.

\begin{edfigure*}[htbp]
    \centering
    \includegraphics[width=1\linewidth]{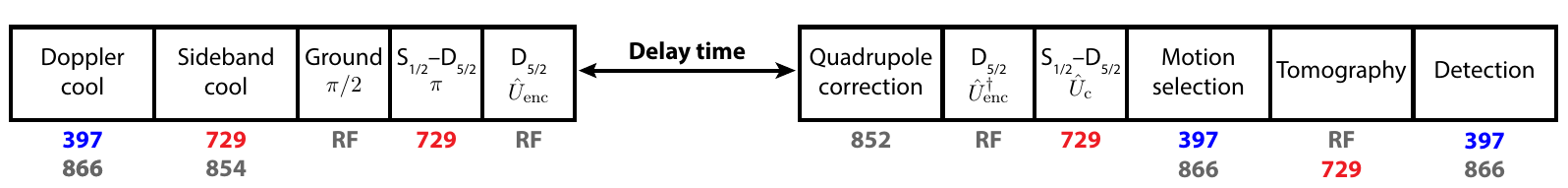}
    \caption{\textbf{Overview of experimental sequence.} A simplified view of the experimental pulse sequence, highlighting which laser or radio-frequency source is used to perform each operation, with time proceeding from left to right. Sideband cooling, $S_{1/2}$ to $D_{5/2}$ $\pi$-pulses, the $\hat{U}_\text{c}$ error-correction operation, and tomography operations contain multiple individual pulses, which are described in detail in the text.  ``Quadrupole correction'' and ``Motion selection'' operations are also described in the text below.}
    \label{fig:pulse-seq}
\end{edfigure*}

\subsubsection{Experimental pulse sequence}

An overview of the experimental pulse sequence is shown in \cref{fig:pulse-seq}, showing which laser or RF sources are used for which operation. The sideband cooling operation is performed by applying a standard sequence of alternating pulses of \SI{729}{\nano\meter} light tuned to the red motional sideband frequency and \SI{854}{\nano\meter} light addressing the $D_{5/2} \leftrightarrow P_{3/2}$, with a total of 36 pulses of each. Other composite operations ($S_{1/2}$ to $D_{5/2}$ $\pi$-pulses, the $\hat{U}_\text{c}$ error-correction operation, and tomography) are described in detail elsewhere in the main text.

\subsubsection{Error correction pulses}

The error correction operation $\hat{U}_\text{c}$ performs a one-way change of the ion's internal state by utilizing the ion's motional mode. This is the same principle as the Cirac–Zoller gate \cite{cirac1995quantum}, and the action of the gate is illustrated for both the presence and the absence of an error in \cref{fig:error-correction-diagram}. When the ion is in the error state $\Ket{\mathcal{E}_1}$, the population is first moved to a ground state superposition via two \SI{729}{\nano\meter} $\pi$-pulses, and is then transferred to the $\Ket{\pm\tfrac52}$ states via a one-way motion adding pulses, putting the ion in a superposition of $\Ket{\pm\tfrac52}\Ket{1}_\text{M}$, provided the ion starts in the motional ground state. If an error is not present, however, it is in a superposition of the $\Ket{\pm\tfrac52}\Ket{0}_\text{M}$ states. Here, the \SI{729}{\nano\meter} carrier $\pi$-pulses coupling $\Ket{D_{5/2};\pm\tfrac32}\Ket{n}_\text{M}\leftrightarrow\Ket{S_{1/2};\pm\tfrac12}\Ket{n}_\text{M}$ have no effect. Then, the blue sideband pulses, which couple $\Ket{S_{1/2};\pm\tfrac12}\Ket{n}_\text{M} \leftrightarrow \Ket{D_{5/2};\pm\tfrac52}\Ket{n+1}_\text{M}$ also have no effect, because there is no $\Ket{S_{1/2};\pm\tfrac12}\Ket{-1}_\text{M}$ state for the ion's $\Ket{D_{5/2};\pm\tfrac52}\Ket{0}_\text{M}$ superposition to couple to, and the ion's internal state remains unchanged.

\begin{edfigure}[htbp]
    \centering
    \includegraphics[width=1\linewidth]{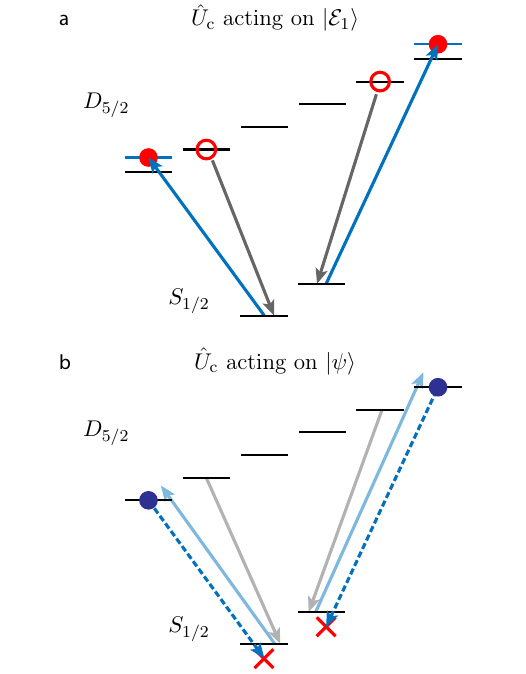}
    \caption{\textbf{Error correction operation in the presence and absence of an error.} Energy level diagrams demonstrating the action of pulses in $\hat{U}_\text{c}$. Energy level spacing is not drawn to scale. \textbf{a,} $\hat{U}_\text{c}$ acting on an error state $\Ket{\mathcal E_1}=\alpha \Ket{-\tfrac32}+\beta\Ket{+\tfrac32}$, with the effect of moving the population from $\Ket{\pm \tfrac{3}{2}} \Ket{0}_\text{M}$ to $\Ket{\pm \tfrac{5}{2}} \Ket{1}_\text{M}$ via blue sideband (motion-adding) pulses. \textbf{b,} $\hat{U}_\text{c}$ acting on an error-free state $\Ket{\psi}=\alpha \Ket{-\tfrac52}+\beta\Ket{+\tfrac52}$, with no effect on the state. Because the ion starts in the $\Ket{0}_\text{M}$ motional state, there is no $\Ket{-1}_\text{M}$ motional state for the sideband pulse to couple $\Ket{\psi}$ to, so there is no change to the ion's state.}
    \label{fig:error-correction-diagram}
\end{edfigure}

\subsubsection{Experimental apparatus}

The \ca~ion is trapped in a cryogenic surface-electrode trap \cite{debry_mbp, bruzewicz2016scalable} loaded from a 2D magneto-optical trap of cold calcium atoms \cite{bruzewicz2019dual}. The \ca~ion is trapped in an RF Paul trap with a \SI{1.9}{\mega\hertz} axial frequency, and the axial motion is cooled via resolved sideband cooling prior to the start of the experimental sequence. The cold stage containing the trap is maintained at \SI{5}{\kelvin}, resulting in long trap lifetimes and low motional heating rates \cite{labaziewicz2008suppression}. The trap geometry and laser beam orientations are illustrated in \cref{fig:trap}.

\begin{edfigure}[htbp]
    \centering
    \includegraphics[width=1\linewidth]{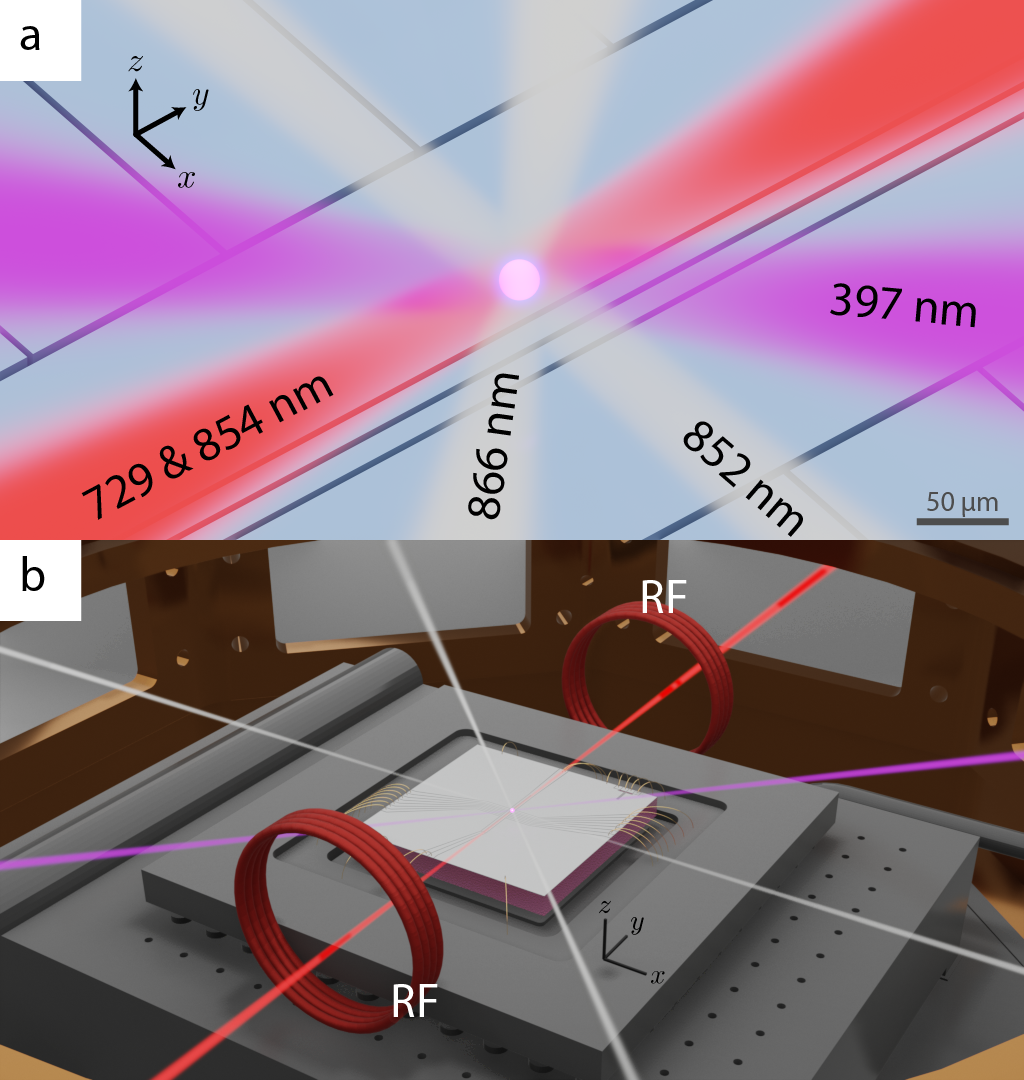}
    \caption{\textbf{Ion trap geometry.} Surface electrode trap and laser beam geometry, including labels of the five beams, and RF antennas, used in the experiment. Ion not drawn to scale. \textbf{a,} Close-up of trapping region with laser beams labeled. \textbf{b,} Rendering of \SI{5}{\kelvin} stage of the cryostat (windows not shown) showing the trap (center) held in a ceramic pin grid array and the locations of the RF antennas that apply all RF pulses during the experiment. The trap (light gray square) is \SI{1}{\centi\meter} on a side.}
    \label{fig:trap}
\end{edfigure}

%%%%%%%%%%
\subsubsection{State preparation}

The initial unencoded qubit state is prepared with a combination of RF pulses addressing M1 transitions and laser pulses addressing E2 transitions. We initialize the ion into the $\Ket{S_{1/2};-1/2}$ state with optical pumping, and then prepare an equal superposition of the $\Ket{S_{1/2};-1/2}$ and $\Ket{S_{1/2};+1/2}$ states using a radiofrequency (RF) $\pi/2$-pulse resonant with the ground state Zeeman splitting (approximately \SI{11}{\mega\hertz}). This superposition is coherently transferred to the $\Ket{-5/2}$ and $\Ket{+5/2}$ states in the $D_{5/2}$ manifold using two optical $\pi$-pulses, creating a spin-cat superposition of $\Ket{\pm J}$ in the $\hat{J_z}$ basis.

The change of basis/logical encoding unitary $U_\text{enc} = e^{i \hat{J}_y \pi/2}$ is then applied via an RF pulse resonant with the $D_{5/2}$ manifold's Zeeman splitting of \SI{6.7}{\mega\hertz} (\cref{fig:circuit}c). This implements an SU(2) rotation of the six level manifold, preparing the logical qubit state $\Ket{\psi_\text{L}}=\alpha \l0 + \beta \l1$ for arbitrary $\alpha,\beta$. These logical states are prepared with a fidelity $\ge0.996(2)$, as shown by tomographic reconstructions of the prepared density matrices in the main text (\cref{fig:denmats}).

%%%%%%%%%%%%%%%%%%%%
\subsubsection{Considerations for extended interrogation times}

The primary use case of the $\hat{J}_z$-based error correction presented here is to preserve the coherence of a memory qubit as it evolves over time in a noisy environment, which requires consideration of additional coherent sources of error. While the scheme protects against small Zeeman frequency detuning errors, the frequency offset cannot be allowed to accumulate such that there are large coherent detuning errors each trial, requiring frequency recalibration. Similarly, the electric and magnetic quadrupole shifts cause coherent errors to the $D_{5/2}$ manifold that can be neglected over very short times but must be counteracted at longer delay times.

While the error-correction scheme does mitigate the effects of a frequency offset, it only works in the small-error regime (like all quantum error correction protocols) and the Zeeman splitting frequency must be know well enough that the average phase picked up on a given trials is close to 0. The magnetic field in this apparatus experiences slow drifts on the timescales of seconds, but this is significantly faster than the time required for a full tomographic reconstruction. To capture realistic performance, the magnetic field must be frequently recalibrated to remain within the correctable range of the code ($\sigma_\phi \ll 1$). We perform a one-shot Ramsey experiment in the ground state qubit with a \SI{2}{\milli\second} interrogation time in between each realization of the error correction experiment to track the drifting magnetic field. After each shot of this experiment, based on the outcome, all pulse frequencies in the experiment are updated with a small frequency shift scaled by each transition's effective $g$-factor, recalibrating the entire pulse sequence without needing to measure each line individually. We extract $\sigma_B = \SI{0.78(7)}{\nano\tesla}$ from the ground state physical qubit's error measurements. This value is used to compute the dephasing parameter $\chi = \frac{1}{2}(\mu_\text{B}\sigma_B/\hbar)^2t^2$ in \cref{fig:breakeven} for each interrogation time $t$.

In addition to slow drifts of the magnetic field on the timescales of seconds or more, the magnetic field noise contains prominent oscillations on timescales on the order of the interrogation time that require calibration as well. We observe significant oscillations in the magnetic field strength at \SI{60}{\hertz} coinciding with the \SI{120}{\volt} line frequency. To prevent this from skewing coherence time measurements, we trigger each experimental realization to begin at the same point in the \SI{60}{\hertz} cycle. This leaves a residual effect where the average magnetic field during an interrogation period changes as a function of the interrogation duration. We measure this shift and include it in the frequency calibration of our pulses for each value of the interrogation time.

Finite interrogation times also reveal a quadrupole shift of the $D_{5/2}$ manifold arising from both the quadratic term of the Zeeman Hamiltonian of $\sim \SI{10}{\hertz}$ (coupling the $D_{5/2} \leftrightarrow D_{3/2}$ manifolds) \cite{UDportal} and the remainder from the electric-field slope along the magnetic field axis $\partial E_z/\partial z$ \cite{roos2006designer}. This leads to a shift proportional to $\hat{J}_z^2$ on the $D_{5/2}$ manifold, increasing the energy of the outer states (with large $|m|$) compared to the inner states. We measure the total quadrupole shift to be \SI{38(1)}{\hertz} between the $|m|=1/2$ and $|m|=5/2$ states. We counteract this by applying an \SI{852}{\nano\meter} laser tuned approximately \SI{1}{\tera\hertz} blue of the \SI{854}{\nano\meter} $P_{3/2} \leftrightarrow D_{5/2}$ in \ca. By polarizing the light along the magnetic field axis ($\pi$-polarized), we can apply an inverse shift proportional to $-\hat{J}_z^2$. We calibrate the shift from this laser against the atomic quadrupole shift and apply a short pulse of \SI{852}{\nano\meter} light $<\SI{1}{\micro\second}$ at the end of each logical qubit delay time to cancel the effects of the $\hat{J}_z^2$ quadrupole shift. The quadrupole shift can also be canceled by a variety of approaches, including setting the magnetic field direction at a particular angle from the trap axes \cite{itano2000external, madej2004absolute} and dynamical decoupling \cite{shaniv2019quadrupoleshiftcancellation}. With this last approach, care must be taken to ensure that the operations are bias-preserving for codes that protect against only $\hat{J}_z$-type errors. Note that our application of a $\hat{J}_z^2$ Hamiltonian via an off-resonant laser is both bias-preserving and straightforward to calibrate, while achieving a precise magnetic field angle can be challenging.

%%%%%%%%%%%%%%%%%%%%
\subsubsection{Density matrix tomography and fidelity measurements}
Measured density matrices are produced using maximum likelihood estimation (MLE) from a tomographically complete set of projective measurements on the $D_{5/2}$ manifold \cite{Hradil2004}. To reconstruct the full density matrix, a minimum of $d^2-1=35$ projectors are needed. In most of the data presented here, a total of 54 projectors are measured, with 200 trials per projector.

The projection operators are rotated versions of the $\hat{J}_z$ basis state projectors $\ket{m_J}\bra{m_J}$. All six are measured from a single shot of the experiment using a qudit-style readout scheme \cite{low2023control, campbell2022polyqubit, ringbauer2022universal}. Each state is probed sequentially by transferring the population to the ground state and then applying \SI{397}{\nano\meter} light to fluoresce any ground state population. This is repeated for the other $D_{5/2}$ levels until fluorescence is observed. The order in which the states are read out is randomized for each trial to minimize the impact of spontaneous decay on the results. Additional projection operators are obtained by rotating the $D_{5/2}$ manifold with combinations of RF $\hat{J}_x$ and $\hat{J}_y$ pulses prior to performing the qudit-style readout. These sets of 6 projection operators are described by the pair of rotation angles for the $\hat{J}_x$ and $\hat{J}_y$ rotations $(\theta_x,\theta_y)$. To collect data on 54 projection operators, measurements were made after performing all combinations of rotations with $\theta_x$ and $\theta_y$ chosen from $\set{0, \pi/4, \pi/2}$, with the $\hat{J}_x$ rotation always applied first.

The density matrix is then estimated by numerically finding the density matrix $\rho_\text{est}$ that minimizes
\begin{align*}
    -\log(\mathcal{L(\rho_\text{est},\,M)}),
\end{align*}
where $\mathcal{L}$ is the likelihood of observing the given set of measurement results $\mathcal{M}$ if the true state was $\rho_\text{est}$. The MLE process is observed to be stable and insensitive to the initial conditions supplied for the test matrix $\rho_\text{est}$. Because the order of measurement of the projection operators is randomized, the effect of measurement errors is to bring the reported density matrix closer to the fully-mixed identity matrix, and the reported fidelities can thus be taken as lower bounds on the true fidelity of the state.

We use the reconstructed density matrices to compute all fidelities reported for the $D_{5/2}$ manifold logical qubit states and unencoded states. We calculate these using $\mathcal{F}(\rho_{\mathrm{est}},\rho_{\mathrm{ideal}})$ (\cref{eq:fidelity-def}) where $\rho_\text{ideal}$ is the desired state's density matrix. Uncertainties on fidelities are obtained via bootstrapping \cite{horowitz_bootstrap} by generating at least 100 \cite{goodhue2012resampling} sets of measurements $\mathcal{M}_i$ with measurement probabilities determined by $\rho_\text{est}$, performing MLE on $\mathcal{M}_i$ to estimate $\rho_{\text{est},i}$, computing fidelity $\mathcal{F}_i = \mathcal{F}(\rho_{\text{est},i},\rho_\text{ideal})$, and calculating the standard deviation of $\{\mathcal{F}_i\}$.

%%%%%%%%%%%%%%%%%%%%
\subsubsection{Procedures employed in error correction evaluation}

Procedures for extracting the fidelity of the error correction scheme, initializing the motional state ancilla for error recovery, and quantifying motional state errors, are described below.

{\bf Error-corrected fidelity measurements.} Fidelities for logical qubits (with and without error correction) are calculated from tomographic data as described above. In the case of the logically-encoded qubit with no error correction, quoted values are taken from a single density matrix measurement, as shown in \cref{fig:correction}b. The error correction sequence has an additional phase that must be calibrated to ensure that the error-free part of the mixed state and the error-corrected part of the mixed state are recombined coherently: the relative phase of the two sideband pulses $\phi_\text{c}$. To extract the fidelity of the error correction scheme, 10 runs of the experiment are performed with different sideband phases $\phi_\text{c}$. For each, the fidelity is calculated from the measured density matrix as shown in \cref{fig:correction}c, and the overall fidelity is the maximum with respect to $\phi_\text{c}$, extracted from a fit of these measurements to a sinusoid (\cref{fig:correction}d). The optimal phase could instead be calculated analytically, if the $S_{1/2} \leftrightarrow D_{5/2}$ laser frequency, motional sideband frequencies, and phases acquired due to light shifts from each previous laser pulse are sufficiently stable and measured accurately. Because the \SI{729}{\nano\meter} laser used in this experiment is observed to drift at a rate of $\sim$\SI{2}{\kilo\hertz} per day, we opted to measure and fit $\phi_\text{c}$ for each data point.

%%%%%%%%%%%%%%%%%%%%
{\bf Motional state initialization.} We perform the autonomous error correction protocol described in the main text by using the motional state of the ion as a fresh ancilla that can be reset. In between each shot of the experiment, a combination of Doppler cooling and resolved sideband cooling is used to initialize the axial mode of the ion's motion to the motional ground state $\Ket{0}_\text{M}$. The ion is trapped \SI{50}{\micro\meter} above the surface of an aluminum-on-sapphire surface electrode trap maintained at \SI{5.5}{\kelvin}. The axial motion has a frequency of \SI{1.9}{\mega\hertz}.

The ion's thermal motion is characterized by measuring the asymmetry of the amplitude of the ion's red and blue motional sidebands following sideband cooling \cite{leibfried2003quantum}. We observe a heating rate of 8.8(5) motional quanta per second (see \cref{fig:heating}).

\begin{edfigure}
    \centering
    \includegraphics[width=\linewidth]{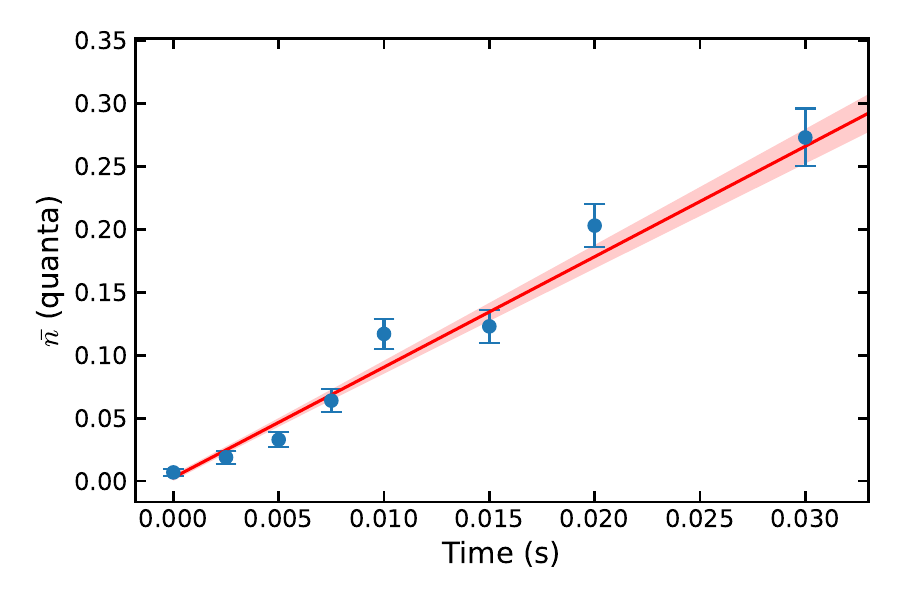}
    \caption{Heating rate measurements and fit performed using sideband thermometry, with a heating rate of \SI{8.8(5)}{\per\second}. Shaded region corresponds to 1$\sigma$ uncertainty on the fit.}
    \label{fig:heating}
\end{edfigure}

% Section on MF QEC alternatives

\begin{edfigure}
    \centering
    \includegraphics[width=1\linewidth]{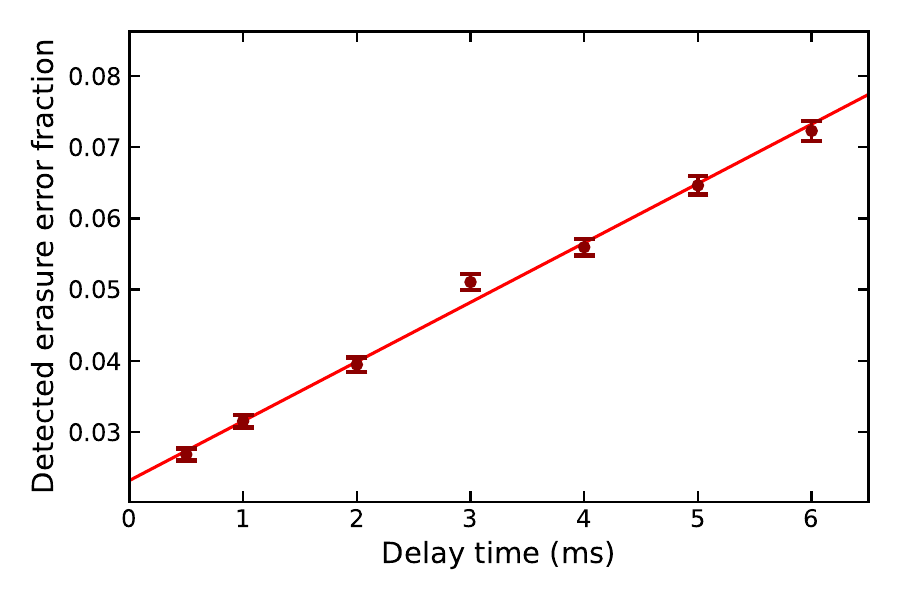}
    \caption{Fraction of trials resulting in erasure detections, due to motional heating, as a function of memory qubit interrogation time. The fit has a slope of \SI{8.3(2)}{\per\second}.}
    \label{fig:erasures}
\end{edfigure}
{\bf Motional state erasure errors in measurement-free QEC.} \label{sec:erasure}
The measurement-free quantum error correction scheme used here relies on the ion being in the motional ground state when $\hat{U}_\text{c}$ is applied, which is technically feasible through sympathetic mid-circuit cooling, but outside the scope of this work. With only a single ion in our experiment, we instead simulate this through a motional-state selective measurement performed near the end of the experimental sequence (\cref{fig:pulse-seq}) and select for trials where the ion was in the ground state. If the ion is not in the motional ground state when the motion-adding sideband pulses are applied, $\hat{U}_\text{c}$ ceases to be a one-way operation and the error-free state $\Ket{\psi}=\alpha \Ket{-\tfrac{5}{2}}+\beta\Ket{+\tfrac{5}{2}}$ is transferred to a ground state superposition. We detect this with the application of fluorescent light on the $S_{1/2} \leftrightarrow P_{1/2}$ cycling transition. Even with a sympathetic coolant ion, we expect this to be useful, as any detections serve as erasure conversions. The threshold for erasure errors is significantly higher than other types of errors, approaching 50\% \cite{stace_erasure_threshold, ohzeki_erasure_threshold}.

These detections are selecting only for thermal excitation of the motional mode---not for $\hat{\mathcal E}_k$ errors---and enable an accurate picture of the performance of this protocol in the absence of motional excitation. The discarded measurements are consistent with the measured motional heating rate of $8.8(5)$ motional quanta per second (\cref{fig:heating}), with a slope of \SI{8.3(2)}{\per\second} (\cref{fig:erasures}). In the case of applied rotation errors with no delay time, we observe no correlation between the strength of the applied errors and the frequency of bright detections. At a delay time of \SI{0}{\milli\second}, we see an approximately 2\% detection rate, consistent with our observed $\approx 99\%$ ground state cooling probability and the $\approx\SI{1}{\milli\second}$ of pulses preceding the motional detection pulse.
%The ion also experiences decay of the $D_{5/2}$ excited states at a rate of $\Gamma \approx \SI{0.8}{\per\second}$, but this is both significantly less than the motional heating rate and also not detected in this protocol due to the \SI{729}{\nano\meter} $\pi$-pulses that would transfer any population that decayed into the $\Ket{\pm\tfrac{3}{2}}$ states before detection.

\subsection{Acknowledgments}
This material is based upon work supported by the National Science Foundation Graduate Research Fellowship under Grant No. 2141064. This research was supported by the U.S. Army Research Office through grant W911NF-24-1-0379. ILC acknowledges support in part by the NSF Center for Ultracold Atoms.  This material is based upon work supported by the Department of Defense under Air Force Contract No. FA8702-15-D-0001. Any opinions, findings, conclusions or recommendations expressed in this material are those of the author(s) and do not necessarily reflect the views of the Department of Defense.

 % \printbibliography
%  \bibliographystyle{apsrev4-2}
% \bibliography{silq}

\putbib
\end{bibunit}

 \end{document}